\documentclass[sigconf, nonacm]{acmart}

\usepackage{balance}       

\usepackage{xcolor}
\usepackage{soul}
\usepackage{subcaption}
\usepackage{float}
\usepackage{comment}
\usepackage{algorithm}
\usepackage{algorithmic}
\usepackage{pgfplots}
\pgfplotsset{compat=1.17}
\usepackage{pgfplotstable}
\usetikzlibrary{matrix}
\usepgfplotslibrary{groupplots}

\newcommand\vldbdoi{XX.XX/XXX.XX}
\newcommand\vldbpages{XXX-XXX}
\newcommand\vldbvolume{14}
\newcommand\vldbissue{1}
\newcommand\vldbyear{2020}
\newcommand\vldbauthors{\authors}
\newcommand\vldbtitle{\shorttitle} 
\newcommand\vldbavailabilityurl{URL_TO_YOUR_ARTIFACTS}
\newcommand\vldbpagestyle{plain} 

\begin{document}
\title{TableDC: Deep Clustering for Tabular Data}

\author{Hafiz Tayyab Rauf} 
\orcid{0000-0002-1515-3187}
\affiliation{%
	\institution{Department of Computer Science, University of Manchester}
	\city{Manchester}
	\country{UK.}
	\postcode{M13 9PL}
}
\email{hafiztayyab.rauf@manchester.ac.uk}
\author{Andre Freitas}
\orcid{0000-0002-4430-4837}
\affiliation{%
	\institution{Department of Computer Science, University of Manchester\\ Manchester, UK.\\ IDIAP Research Institute\\ Martigny, Switzerland.}
	\postcode{M13 9PL}
}
\email{andre.freitas@manchester.ac.uk}

\author{Norman W. Paton}
\orcid{0000-0003-2008-6617}
\affiliation{%
	\institution{Department of Computer Science, University of Manchester}
	\city{Manchester}
	\country{UK,}
	\postcode{M13 9PL}
}
\email{norman.paton@manchester.ac.uk}

\begin{abstract}
Deep clustering (DC), a fusion of deep representation learning and clustering, has recently demonstrated positive results in data science, particularly text processing and computer vision. However, joint optimization of feature learning and data distribution in the multi-dimensional space is domain-specific, so existing DC methods struggle to generalize to other application domains (such as data integration and cleaning). In data management tasks, where high-density embeddings and overlapping clusters dominate, a data management-specific DC algorithm should be able to interact better with the data properties for supporting data cleaning and integration tasks. This paper presents a deep clustering algorithm for tabular data (TableDC) that reflects the properties of data management applications, particularly schema inference, entity resolution, and domain discovery. To address overlapping clusters, TableDC integrates Mahalanobis distance, which considers variance and correlation within the data, offering a similarity method suitable for tables, rows, or columns in high-dimensional latent spaces. TableDC provides flexibility for the final clustering assignment and shows higher tolerance to outliers through its heavy-tailed Cauchy distribution as the similarity kernel. The proposed similarity measure is particularly beneficial where the embeddings of raw data are densely packed and exhibit high degrees of overlap.  
Data cleaning tasks may involve a large number of clusters, which affects the scalability of existing DC methods. TableDC's self-supervised module efficiently learns data embeddings with a large number of clusters compared to existing benchmarks, which scale in quadratic time. We evaluated TableDC with several existing DC, Standard Clustering (SC), and state-of-the-art bespoke methods over benchmark datasets. TableDC consistently outperforms existing DC, SC, and bespoke methods. 
\end{abstract}
\maketitle

\pagestyle{\vldbpagestyle}
\begingroup\small\noindent\raggedright\textbf{PVLDB Reference Format:}\\
\vldbauthors. \vldbtitle. PVLDB, \vldbvolume(\vldbissue): \vldbpages, \vldbyear.\\
\href{https://doi.org/\vldbdoi}{doi:\vldbdoi}
\endgroup
\begingroup
\renewcommand\thefootnote{}\footnote{\noindent
This work is licensed under the Creative Commons BY-NC-ND 4.0 International License. Visit \url{https://creativecommons.org/licenses/by-nc-nd/4.0/} to view a copy of this license. For any use beyond those covered by this license, obtain permission by emailing \href{mailto:info@vldb.org}{info@vldb.org}. Copyright is held by the owner/author(s). Publication rights licensed to the VLDB Endowment. \\
\raggedright Proceedings of the VLDB Endowment, Vol. \vldbvolume, No. \vldbissue\ %
ISSN 2150-8097. \\
\href{https://doi.org/\vldbdoi}{doi:\vldbdoi} \\
}\addtocounter{footnote}{-1}\endgroup

\ifdefempty{\vldbavailabilityurl}{}{
\vspace{.3cm}
\begingroup\small\noindent\raggedright\textbf{PVLDB Artifact Availability:}\\
The source code, data, and/or other artifacts have been made available at \url{https://github.com/hafizrauf/TableDC}.
\endgroup
}

\section{Introduction}

Deep Clustering (DC) combines the unsupervised grouping of related data items with the learning of clustering-friendly representations for the data to be clustered. DC uses deep neural network architectures with unsupervised clustering mechanisms to effectively classify multi-dimensional, unlabeled data. Early implementations primarily used autoencoders for dimensionality reduction, where the embeddings in the latent space were clustered using standard clustering (SC) algorithms such as K-means \cite{DBLP:conf/icml/YangFSH17}. Advancements in DC have led to more sophisticated architectures, based on variational autoencoders \cite{DBLP:journals/corr/DilokthanakulMG16}, Generative Adversarial Networks \cite{DBLP:conf/aaai/MukherjeeALK19}, and self-supervised learning paradigms \cite{DBLP:conf/nips/Alwassel0KTGT20}. Recent DC methods focus on integrating representation learning and cluster assignment within end-to-end trainable models using graph-based neural networks \cite{DBLP:conf/aaai/LiuTZLSYZ22, DBLP:conf/aaai/TuZL0CZC21, DBLP:conf/www/Bo0SZL020} to capture relational structures within data. 



DC has been employed successfully in several domains, particularly image processing~\cite{DBLP:journals/corr/abs-2206-07579, DBLP:journals/corr/abs-2206-07579,DBLP:journals/corr/abs-2311-14310}, which includes downstream tasks such as segmentation and object detection \cite{DBLP:journals/corr/abs-2206-07579}, clustering of in and out-of-distribution noise in corrupted images \cite{DBLP:conf/eccv/AlbertAOM22}, and unsupervised image alignment and clustering \cite{DBLP:conf/cvpr/JinWDLZ23,DBLP:journals/corr/abs-2311-14310}. 
Further typical applications of DC include community and anomaly detection \cite{DBLP:journals/wpc/EmadiM18, DBLP:conf/ijcai/LiuX0ZHPNYY20} and medical applications \cite{DBLP:conf/miccai/KartBGR21}. These DC proposals excel in determining complex patterns within images and graphs, extracting features, and subsequently clustering similar images and communities based on those patterns. The successful application of DC in specific domains has led to the development of more specialized algorithms that reflect specific properties of the domain features, even if the algorithms are not explicitly domain-specific. For example, in \cite{DBLP:conf/cvpr/ChenCPLW22}, the authors use domain-specific 3D geometric consistency to guide the learning of 2D image representation and solve several 2D-image-based downstream tasks. 


In data management, and in particular data integration and cleaning, a variety of important problems make use of clustering, including {\it schema inference}~\cite{DBLP:conf/er/Kellou-MenouerK15,DBLP:conf/edbt/BonifatiDM22}, {\it entity resolution} \cite{DBLP:journals/is/PapadakisMGSTGB20,DBLP:journals/datamine/CostaMO10,DBLP:journals/pvldb/HassanzadehCML09} and {\it domain discovery} \cite{DBLP:journals/vldb/PiaiAMS23,DBLP:journals/pvldb/OtaMFS20}.  We have investigated the application of existing DC algorithms to these three problems, with positive results; DC algorithms consistently outperform established SC algorithms for a variety of initial data representations for all of tables, columns and rows \cite{DBLP:journals/corr/abs-2305-13494}. However, existing DC algorithms do not fully exhibit the  properties required to address the clustering of latent representations of table objects, required to support the transfer to data cleaning and integration tasks.  In our previous work \cite{DBLP:journals/corr/abs-2305-13494}, we observed that existing DC algorithms often fall short when applied to data integration tasks, where the nature of the data is fundamentally different from that in image processing applications, with tables, rows or columns being compared in the latent space instead of images \cite{DBLP:journals/corr/abs-2305-13494}.

In this paper, we propose a new deep clustering algorithm for tabular data (TableDC) that reflects the properties of table embeddings and data management applications, specifically: (i) Data representations are typically dense, where features are closely packed together. As the dimensionality increases, it becomes more challenging for the models to interpret the relationships between features. High density in data reduces the differences in features and makes it difficult to differentiate between clusters effectively. The closeness of data points in the latent space leads to ambiguity, where a data point could reasonably belong to multiple clusters. (ii) Tabular data contains higher semantic and syntactic variability levels than image data, leading to increased ambiguity and cluster overlap. Tables, columns and rows commonly involve the composition of different types of semantic objects (e.g., named entities, categories, numbers), each in their own way subject to ambiguity, leading to potentially overlapping embeddings when projected to high-dimensional latent spaces. In contrast, while complex, image data is based on a single data modality, with the expectation of smoother transfer of visual features across domains, naturally conforming to a better cluster separation. (iii) In a data management setting, data can have a large number of clusters compared to existing DC applications. In our previous work \cite{DBLP:journals/corr/abs-2305-13494}, we observed that existing DC algorithms struggle with scalability when the data has a large number of clusters.  This affects the DC method's performance in terms of the high probability of misclassification, a higher computational burden, and cluster instability, where small changes in data and model parameters lead to significant changes in cluster assignments.

The contributions of this paper are:
\begin{enumerate}
    \item The identification of features of data management applications that challenge the assumptions that underpin existing DC algorithms.
    \item The proposal of a new DC algorithm, TableDC, that builds on the analysis at (1) to provide an algorithm that: (i) Maintains a balance between cluster compactness and separation during cluster center initialization 
    to better assist TableDC training with robust cluster shapes. (ii) Integrates Mahalanobis distance, a covariance-aware distance that considers the data's variance and covariance to handle dense features, unlike Euclidean distance, which treats all dimensions equally. 
    (iii) Offers flexibility in overlapping clusters through its heavy-tailed Cauchy distribution as a similarity kernel, especially when the cluster boundaries are ambiguous. (iv) Scales in quasi-linear time under significant increases in the number of clusters. 
    \item The experimental evaluation of TableDC from (2) with both state-of-the-art SC and DC algorithms and recent bespoke algorithms on real-world datasets for {\it schema inference}, {\it entity resolution} and {\it domain discovery}, which shows that TableDC significantly outperforms existing techniques.
\end{enumerate}



\section{Related Work}
\label{sec:related}

This section reviews related work in areas of relevance to the development of a deep clustering algorithm for data management applications, specifically: (i) {\it existing DC algorithms, their features and applications} 
; and (ii) {\it clustering in data management}, focusing on the three applications on which we evaluate techniques and representations for tables, columns and rows.

\subsection{Existing DC algorithms}

The basic DC architecture consists of two phases: (i) representation learning and (ii) clustering. The joint optimization of both phases helps each to improve the algorithm's performance iteratively. A precursor to (i) and (ii) is (iii) cluster initialization, which contributes to the algorithm's performance by providing the initial data shape. With better cluster initialization, a DC algorithm is less likely to get stuck in local minima, and requires fewer converging iterations.

Regarding (i), the most widely used architecture is the basic Auto-encoder (AE) \cite{DBLP:conf/icml/YangFSH17}, where the encoder function \( f_{e} \) encodes an input representation \( x_{i} \) into a latent representation given by \( h_{i} = f_{e}(x_{i}) = \frac{1}{1 + e^{-(Wx_{i} + b_{i})}} \). Similarly, the decoder function \( f_{d} \)  decodes \( h_{i} \) back into the reconstructed input \( x'_{i} = f_{d}(h_{i}) \). Here, \( W \) and \( b_{i} \) represent the weights and biases of the neural network, respectively. The optimization function of AE with \( N \) samples is defined as \( f_{min} = \min \frac{1}{N} \sum_{i=1}^{N} \|x_{i} - x'_{i} \|^2 \). Many recent DC proposals use generative models in representation learning, especially Variational Auto Encoders (VAEs), which combine an AE architecture with probabilistic graphical models.  In VAEs, the encoder maps the input data into a probabilistic distribution over the latent space while the decoder reconstructs the data from the latent probabilistic representation. 
Some recent proposals with VAEs in representation learning are DGG \cite{DBLP:conf/iccv/YangCL019}, LTVAE \cite{DBLP:conf/iclr/LiCPZ19}, VaDE \cite{DBLP:conf/ijcai/JiangZTTZ17} and MFCVAE \cite{DBLP:conf/nips/FalckZWNYH21}. 

To benefit from the local and global structure of data in representation learning, several recent DC proposals combine basic AE learning with  a Graph Convolutional Network (GCN) to assist the model in learning the structural information. 
Some recent examples are DCRN \cite{DBLP:conf/aaai/LiuTZLSYZ22}, DFCN \cite{DBLP:conf/aaai/TuZL0CZC21} and SDCN \cite{DBLP:conf/www/Bo0SZL020}. Subspace representation learning is another recent technique that identifies subspaces that optimize cluster separation, unlike AEs and VAEs, which require additional clustering mechanisms. Subspace learning seeks to identify and represent underlying subspaces within high-dimensional data, while AEs and VAEs focus on reconstructing the input data as accurately as possible. The latest DC methods based on Subspace representation learning are EDESC \cite{DBLP:conf/cvpr/CaiFGWZ022}, DMCAG \cite{DBLP:conf/ijcai/Cui0PP023}, SAGSC \cite{DBLP:conf/aaai/FettalLN23} and OMSC \cite{DBLP:conf/kdd/ChenW0LY22}.

DC is inherently more complex for the tabular data embeddings obtained by Large Language Models (LLMs) than widely used sparse image or textual datasets like CIFAR-10 \cite{krizhevsky2009learning}, CIFAR-100 \cite{krizhevsky2009learning}, STL-10 \cite{DBLP:journals/jmlr/CoatesNL11}, ImageNet \cite{DBLP:journals/ijcv/RussakovskyDSKS15}, USPS \cite{DBLP:conf/icpr/LeCunMBDHHHJB90} and Reuters \cite{DBLP:journals/jmlr/LewisYRL04}, primarily due to data structure and feature representation differences. Image datasets include spatial and visual coherence that DC methods can better differentiate, even when the resolution is low or the number of classes is high.
DC has been used widely on complex and varied datasets unlike ImageNet, where, despite the vast diversity, the visual patterns (such as extracted from different textures, shapes, objects, and backgrounds)
retain a level of consistency 

On the other hand, the primary issue with data management datasets is the high density and significant overlap in the feature space of their embeddings; specifically, tabular data embeddings represent a complex relationship among different features, including categorical and numerical data.


In relation to (ii), clustering, the aim of existing work on DC is to use the low dimensional representation from the representation learning module and obtain clustering assignment probabilities for soft clustering. In the related work, several clustering mechanisms are employed to jointly work with representation learning for optimal clustering. The most widely used one is self-supervised clustering \cite{DBLP:conf/iccv/DizajiHDCH17, DBLP:conf/ijcai/GuoGLY17, DBLP:conf/aaai/PengFLYY17,DBLP:conf/icml/XieGF16, DBLP:conf/www/Bo0SZL020, DBLP:conf/aaai/TuZL0CZC21}, which aims to group data points based on learned representations without explicit labels; the approach uses two distributions: the cluster assignment distribution \(Q\) representing the probability that a data instance belongs to a particular cluster, and an auxiliary distribution \(P\) which is a modified version of \(Q\)  that emphasizes more confident cluster assignments and minimizing the Kullback-Leibler (KL) divergence between \(Q\) and \(P\). \(Q\) is calculated using the distance between the data point \( x_i \) and each cluster centroid \( \mu_k \). The closer \( x_i \) is to \( \mu_k \), the higher the probability  \(Q\).  Self-supervised clustering focuses on high-confidence instances due to the squaring of probabilities in \(P\) 
and preventing cluster dominance by normalizing the probabilities, ensuring balanced clustering assignments.

Pseudo-labeling is another clustering technique recently utilized in DC \cite{DBLP:journals/tip/NiuSW22,DBLP:journals/corr/abs-2206-07579}, where the model generates labels 
that are then used to fine-tune the model, effectively creating a feedback loop where the model iteratively refines its clustering assignment. The recent success of contrastive learning has led to its use in DC \cite{DBLP:conf/ijcai/SunWYPY23, DBLP:conf/ijcai/ZhangYW23, DBLP:conf/aaai/LiuYZLW0TLDC23, DBLP:conf/ijcai/MaK22}, where the data points from low dimensional representations are paired in a way that the same cluster (positive pairs) is contrasted with the different clusters (negative pairs). Further improvements in contrastive deep clustering include the contrast comparison between instance-to-instance, instance-to-cluster, and cluster-to-cluster \cite{DBLP:journals/corr/abs-2206-07579}.

TableDC employs a self-supervised clustering approach by comparing the encodings of tables, rows or columns within a latent space. The selection of an appropriate distance function is critical in this context, as it determines how the distances between each table, row or column and the cluster centroids in the latent space are measured. A simple distance metric like Euclidean distance is beneficial when dealing with data where features exhibit limited variance and correlation, as in image vectors or sparse data matrices \cite{DBLP:conf/www/Bo0SZL020, DBLP:conf/aaai/TuZL0CZC21}. However, a more sophisticated approach is required in dense embeddings, where the data points show significant feature correlation. Here, the Mahalanobis distance is particularly advantageous \cite{mahalanobis2018generalized}. Its ability to account for feature correlation enables the model to discern even subtle differences among clusters, mainly when a cluster shows high similarity across multiple features. Mahalanobis distance is particularly relevant when comparing tables, columns or rows, as it allows for a more nuanced interpretation and grouping of data points based on their inherent relationships and similarities.

In relation to (iii), most existing works use K-means to initialize clusters, whether self-supervised or subspace clustering such as EDESC \cite{DBLP:conf/cvpr/CaiFGWZ022}, DCRN \cite{DBLP:conf/aaai/LiuTZLSYZ22}, DFCN \cite{DBLP:conf/aaai/TuZL0CZC21} and SDCN \cite{DBLP:conf/www/Bo0SZL020}. The quality of the initial clusters significantly affects the final clusters. 
To our knowledge, there is no related work on different cluster initialization approaches in the DC environment. To address this, we compare different cluster initialization approaches in an ablation study in Section \ref{app:cluster_initialization}.

\subsection{Clustering in data management}

Clustering is important in data management for several data integration and cleaning tasks, including those discussed here. 

\textit{Schema Inference} identifies the raw data's types, relationships and attributes to infer an overarching schema \cite{DBLP:journals/vldb/Kellou-MenouerK22}. Schema inference has been explored to infer a schema that fully describes some closely related datasets ~\cite{DBLP:journals/vldb/BaaziziCGS19,DBLP:conf/vldb/BexNV07} or to summarise structures that appear in more diverse datasets. A common first step in the latter case is the identification of recurring structural patterns, which can be used as types in an inferred schema. In this setting, a variety of SC algorithms have been used to group similar structures (e.g., \cite{DBLP:conf/edbt/BonifatiDM22,DBLP:conf/er/Kellou-MenouerK15,DBLP:conf/doceng/TsuboiS19}).


\textit{Entity Resolution} is the well-explored problem of identifying two or more records that represent the same real-world object \cite{DBLP:journals/csur/ChristophidesEP21}. Several proposals use clustering, especially for entity matching and duplicate detection, for which empirical comparisons have been carried out~\cite{DBLP:journals/pvldb/HassanzadehCML09,DBLP:journals/datamine/CostaMO10, DBLP:conf/esws/SaeediPR18,DBLP:journals/jdiq/DraisbachCN20,DBLP:journals/pvldb/HassanzadehCML09}. Although many entity resolution proposals focus on pairwise comparisons, some proposals  incorporate clustering because the transitive closure of pairwise similarity may not always lead to robust clusters \cite{DBLP:conf/kdd/FisherCWR15,DBLP:journals/is/Mandilaras0GSTG21}.

\textit{Domain Discovery} involves identifying columns that draw values from the same collection of values. Most existing works have used bespoke algorithms \cite{DBLP:conf/kdd/LiHG17,DBLP:journals/pvldb/OtaMFS20, DBLP:journals/vldb/PiaiAMS23}. However, we cast domain discovery as a clustering problem, where the goal is to cluster a set of columns that share semantic types. RaF-STD \cite{DBLP:journals/vldb/PiaiAMS23} uses clustering to discover semantic types for heterogeneous sources. Similarly, D4 \cite{DBLP:journals/pvldb/OtaMFS20} provides a column-based clustering approach to discovering local and strong domains from a set of columns, while handling the challenge of incomplete columns. Building on language models, Starmie \cite{DBLP:journals/pvldb/FanWLZM23} provides column clustering based on the learned column representation through contrastive learning.



\section{Proposed Model}
\label{sec:algorithm}


We propose a clustering algorithm, TableDC, that benefits from representation learning in an end-to-end framework to perform several tabular data integration tasks, particularly schema inference, entity resolution, and domain discovery. TableDC considers features of embeddings explicitly in the latent space during its training, such as preserving correlation among highly dense data features through the Mahalanobis distance measure and using a heavy-tailed similarity kernel that provides the flexibility of cluster assignments when the cluster boundaries are ambiguous.

 The representation learning architecture of TableDC is based on self-supervised learning \cite{DBLP:conf/www/Bo0SZL020}. 
 The self-supervised module adopts Mahalanobis distance \cite{mahalanobis2018generalized} between data points and their centroids as it naturally accounts for the variance of different dimensions and is less sensitive to noise, ensuring that no particular feature influences the distance measure due to its scale. Unlike learning a Euclidean distance in a self-supervised module \cite{DBLP:conf/www/Bo0SZL020}, which assumes that all dimensions of the data are orthogonal for the clustering, the Mahalanobis distance shows some degree of correlation between different dimensions by weighting the importance of all dimensions, which could be different based on their semantics.
	
\begin{figure*}[t]
  \centering
  \includegraphics[width=0.9\linewidth,keepaspectratio]{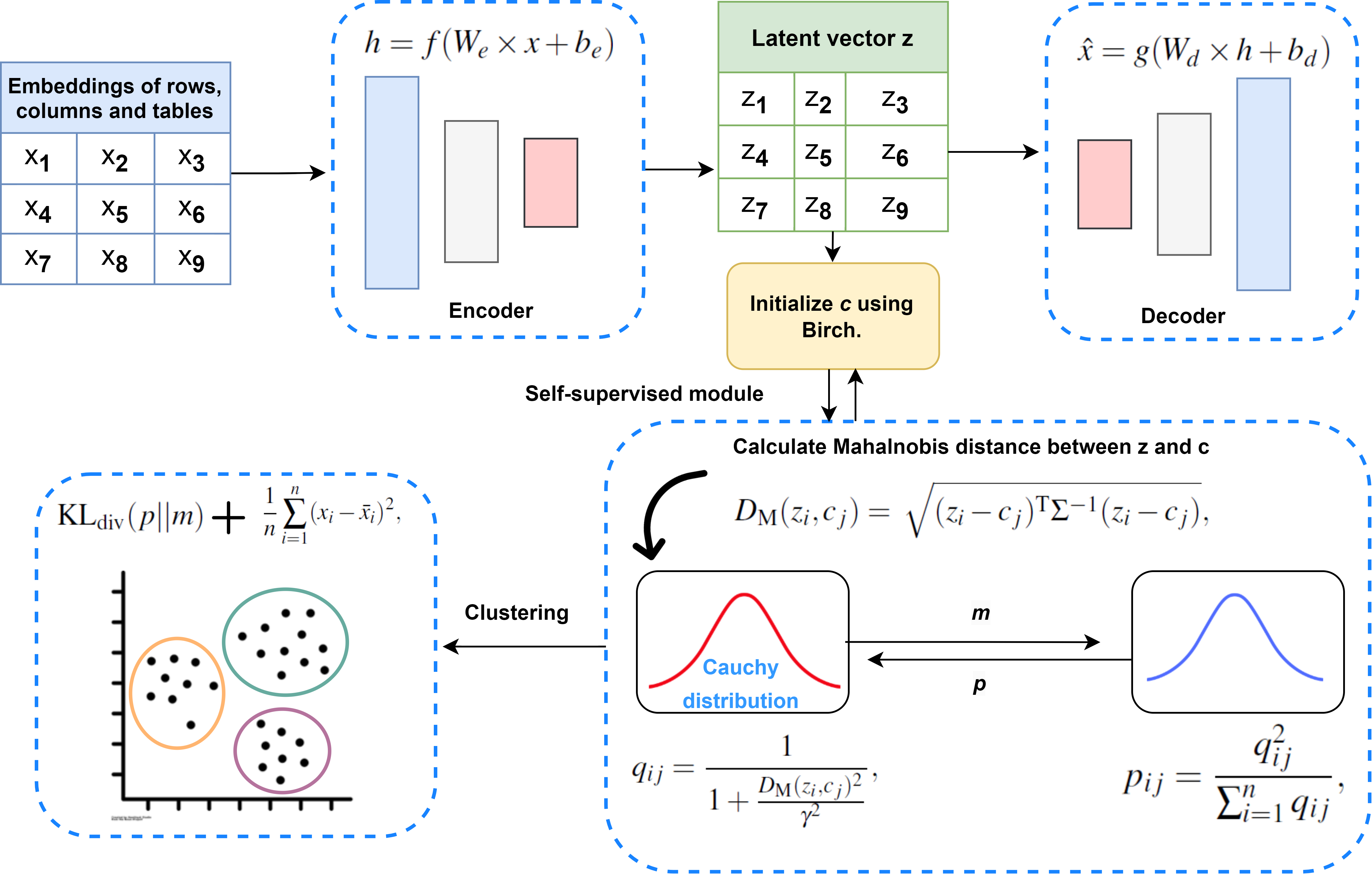}
  \caption{Overview of TableDC framework. An autoencoder takes an embedding matrix as input, representing a set of rows for entity resolution, a set of columns for domain discovery, or a set of tables for schema inference. $\textit{z}$ is the latent representation obtained from the autoencoder, and $\textit{c}$ is initialized using Birch. In the self-supervised module, the $\textit{m}$ distribution is calculated from $\textit{q}$ by taking the Mahalanobis distance between $\textit{z}$ and $\textit{c}$ and setting the Cauchy distribution as a kernel to measure the similarity. $\textit{p}$ is the target distribution, which is also calculated from $\textit{q}$.}
  \label{fig:Expframwork}
\end{figure*}

TableDC integrates an autoencoder for representation learning and a self-supervised module to optimize representation for clustering. An autoencoder consists of two main parts: the encoder and the decoder. The encoder function compresses the input $x$ (an embedding matrix of tables, rows or columns) into a latent-space representation. Mathematically, it can be represented as: 
\begin{equation} \label{eq1}
	h = f(W_e \times x + b_e)
\end{equation}
where $W_e$ is the weight matrix, $b_e$ is the bias term, and $f$ is an activation function for the encoder, such as sigmoid or ReLU \cite{DBLP:conf/icml/NairH10}. The result, $h$, is the encoded representation of the input $x$. The decoder function works to reconstruct  $x$ from the internal representation. It aims to map the encoded data back from the latent space to the original space. The decoder can be represented as:
\begin{equation} \label{eq2}
	\hat{x} = g(W_d \times h + b_d)
\end{equation}
where $h$ is the encoded representation from the encoder, $W_d$ is the weight matrix, $b_d$ is the bias term, and $g$ is an activation function for the decoder. The result, $\hat{x}$, is the reconstructed representation of the original input $x$. An overview of the TableDC framework is presented in Figure \ref{fig:Expframwork}.

Assume we have a latent vector with $n$ data points $z = {z_1, z_2, \dots, z_n}$, where each $z$ represents a row, column or table depending on the data integration problem. We want to cluster the data points into $\mathbb{K}$ clusters with centers $c = {c_1, c_2, \dots, c_k}$. We create a covariance matrix $\Sigma$ to compute the Mahalanobis distance between the data points $z = {z_1, z_2, \dots, z_n}$ and the cluster centers $c = {c_1, c_2, \dots, c_k}$. Since we have noisy data (for example, columns with missing instances, similar tables with different unit measurements, and duplicate rows that affect the embeddings), the variance between different dimensions misleads the importance of the features.  We aim to minimize the noise effect when calculating the covariance by scaling the identity matrix. The scaled identity matrix ensures that all features do not have the same importance to avoid the Euclidean effect and also handles multicollinearity in the data when the covariance matrix is singular. In the case of overlapping clusters, it avoids noisy correlations. The covariance matrix is created by scaling the identity matrix with a scale factor $ \delta = 0.01$ to adjust the strictness of distance between data points:
\begin{equation} \label{eq3}
	\Sigma = \delta \cdot I,
\end{equation}
where $I$ is the identity matrix. We calculate the Cholesky decomposition of the covariance matrix $\Sigma$ to  simplify its inversion. The Cholesky decomposition is a decomposition of a Hermitian, positive-definite matrix into the product of a lower triangular matrix $L$ and its conjugate transpose $L^T$. In our case, since the covariance matrix $\Sigma$ is real and symmetric, the Cholesky decomposition can be represented as:

\begin{equation} \label{eq4}
	C = LL^T,
\end{equation}
\noindent
where $C$ represents the Cholesky decomposition of the covariance matrix $\Sigma$, $L$ is the lower triangular matrix with real and positive diagonal entries, and $L^T$ is the transpose of $L$. The matrix $L$ is computed using the Cholesky decomposition algorithm.

The Cholesky decomposition is used to solve a linear system to obtain the Mahalanobis distance, as it provides an efficient way to compute the inverse of the covariance matrix:

\begin{equation} \label{eq5}
	\Sigma^{-1} = (C)^{-1} = L^{-T} L^{-1}.
\end{equation}

By using the Cholesky decomposition, we can efficiently solve the linear system required to compute the Mahalanobis distance between the latent representations and the cluster centers.

\begin{equation} \label{eq6}
	D_\text{M} (z_i, c_j) = \sqrt{(z_i - c_j)^\text{T} \Sigma^{-1} (z_i - c_j)},
\end{equation}
where $\Sigma$ is the covariance matrix.

The Student’s t-distribution \cite{van2008visualizing} is a common choice for existing DC applications as a kernel to measure the similarity between data points and their cluster centers. However, due to its large degree of freedom parameter, the Student’s t-distribution shows less tolerance to outliers and overlapping clusters; it becomes closer to a normal distribution. In data integration, where the data is more prone to overlaps, i.e., the same magnitude with different semantics, we integrate the Cauchy distribution \cite{cauchy1847methode} with undefined mean and variance, making it robust to outliers, as its shape is unaffected by them. We calculate the soft assignments $q$ for each data point to each cluster using the \textit{Cauchy distribution}. Soft assignments $q$ can be calculated using the Mahalanobis distance:
\begin{equation} \label{eq7}
	q_{ij} = \frac{1}{1 + \frac{D_\text{M} (z_i, c_j)^2}{\gamma^2}},
\end{equation}
where $\gamma$ is a hyperparameter for the Cauchy distribution. We then normalize $q$ to ensure that the sum of the soft assignments for each data point is 1:
\begin{equation} \label{eq8}
	q = \frac{q}{\sum_{j=1}^k q_{ij} + \epsilon},
\end{equation}
where $\epsilon$ is a small value added to prevent division by zero. We then compute the predicted probabilities by applying the softmax function to $q$:
\begin{equation} \label{eq9}
	m = \frac{e^q}{\sum_{j=1}^k e^{q_{ij}}}.
\end{equation}
\subsection{Loss Function}
\label{loss_function_3.1}
Since $m$ acts as clustering probabilities and can be further optimized, we integrate KL-divergence to refine the clustering probabilities $m$ to be more accurate and representative of the underlying data. KL divergence provides a flexible way to guide the model toward the expected clustering assignments. We use the following objective function to optimize the $m$.
\begin{equation} \label{eq10}
	ce_{loss} = \text{KL}_\text{div} (p || m) = \sum_{i=1}^n \sum_{j=1}^k p_{ij} \cdot \log \frac{p_{ij}}{m_{ij}},
\end{equation}
\begin{equation}\label{eq11}
	p_{ij} = \frac{q_{ij}^2}{\sum_{i=1}^n q_{ij}},
\end{equation}
$ce_{loss}$ is clustering loss, where $p_{ij}$ is the target distribution, calculated by adjusting the soft assignments $q$.  This step aims to make the soft assignments more robust during the clustering process. We use the following reconstruction loss ($re_{loss}$) to maintain the inherent structure and essential features in a compressed form.
\begin{equation}\label{eq12}
	re_{loss} = \frac{1}{n} \sum_{i=1}^n (x_i - \bar{x}_i)^2,
\end{equation}
where $\bar{x}$ is the reconstructed input.       $ce_{loss}$ and $re_{loss}$ work together to achieve effective clustering while maintaining the quality of the data reconstruction. The total loss is given by:
\begin{equation} \label{eq13}
	loss = \alpha \cdot ce_{loss} + re_{loss},
\end{equation}
where $\alpha$ is a weighting factor that balances the contribution of the $ce_{loss}$ and the reconstruction loss ($re_{loss}$) to the total loss. Combining $ce_{loss}$ and $re_{loss}$ enables autoencoder learning to generate meaningful latent representations that can be effectively clustered while preserving the structure and information in the input data. We show the pseudo-code for training of TableDC in Algorithm \ref{algo1}.

\begin{algorithm}[t]
\caption{Pseudo-code for training of TableDC}
\label{algo1}
\begin{algorithmic}[1] 
\REQUIRE $x$: representation of rows, columns or tables; $\mathbb{K}$: number of clusters; $maxiter$: maximum number of iterations
\ENSURE $\textit{m}$: clustering probabilities
\STATE Initialize $W_e, b_e, W_d, b_d$ by pre-training autoencoder
\STATE Initialize cluster centers $c$ using \textit{Birch} algorithm
\FOR{$i = 1$ to $\text{length}(maxiter)$}
        \STATE Generate latent representation $z$ from autoencoder using (\ref{eq1})
        \STATE  Compute inverse of covariance matrix $\Sigma^{-1}$ using Cholesky decomposition from (\ref{eq5})
        \STATE  Calculate Mahalanobis distance $D_\text{M}$ between $z$ and $c$ using (\ref{eq6})
        \STATE Apply Cauchy distribution on $D_\text{M}$ to get soft assignments $q$ from (\ref{eq7})
        \STATE Normalize $q$ and apply softmax to get predicted probabilities $m$
        \STATE Calculate target distribution $p$ using (\ref{eq11})
        \STATE Calculate $ce_{loss}$ and $re_{loss}$ from (\ref{eq10}) and (\ref{eq12}) respectively
        \STATE Back propagate and update parameters in TableDC      
\ENDFOR
\RETURN Updated $m$ 
\end{algorithmic}
\end{algorithm}


The reconstruction loss, $re_{loss}$ (Equation \ref{eq12}), of TableDC is consistently reducing over epochs during the training (see ablation study in Section \ref{app:loss_function}), which shows that TableDC is learning a meaningful representation of the data, capturing essential features and structures. In $ce_{loss}$ (Equation \ref{eq10}), $\textit{m}$ is a softmax-ed $\textit{q}$ consistently diverging from the target distribution $\textit{p}$ (see ablation study in Section \ref{app:loss_function}). After applying softmax, $\textit{q}$ becomes a probability distribution $\textit{m}$ over clusters. Each row of the probability matrix has a sum of 1, representing the probability for each point to belong to each cluster. The softmax function emphasizes the significant values in a probability matrix  $\textit{m}$ while suppressing smaller values, making TableDC soft assignments sharper. The choice of the hyperparameter $\alpha$ allows for fine-tuning the trade-off between clustering performance and reconstruction quality. We use $\alpha$ = 0.9 (Equation \ref{eq13}), so TableDC gives more importance to the clustering loss, $ce_{loss}$ (Equation \ref{eq10}) than $re_{loss}$. We determined the weight of $\alpha = 0.9$  for the $ce_{loss}$ through empirical evaluation. 

The presence of noise or exact probabilities towards more than one cluster can help provoke the model towards more confident cluster assignments despite having inherent ambiguities. $\textit{p}$ is artificially derived from $\textit{q}$ by re-normalizing and making high-probability assignments higher and diminishing low-probability assignments. However, the high divergence of $\textit{m}$ from the sharped $\textit{p}$ shows that TableDC is more confident with the actual soft assignments $\textit{q}$ becoming sharper over time. Moreover, the data points in target $\textit{p}$ need more intra-cluster distance instead of high cohesion. A consistent increase in divergence of $\textit{m}$ from $\textit{q}$ indicates that TableDC's internal clustering, as represented by $\textit{m}$, is moving in a direction that better aligns with the true data structure. $\textit{p}$ is an auxiliary target to guide the clustering process and does not represent the ground truth; in the presence of an overlapping cluster, the divergence from $\textit{p}$ is not inherently poor. 
Considering this phenomenon, we can consider the KL-divergence as a maximization problem where we prefer $\textit{m}$ to have a maximum divergence from $\textit{p}$ and can be defined as a sequence of KL-divergences over \( N \) iterations: $
\{D_{KL}(p \parallel m^1), D_{KL}(p \parallel m^2), \ldots, D_{KL}(p \parallel m^N)\}$ where \( D_{KL}(p \parallel m^k) \) represents the KL-divergence between \( p \) and \( m \) at the \( k^{th} \) iteration. The KL divergence is increases over  \( N \) if: $D_{KL}(p \parallel m^{k+1}) > D_{KL}(p \parallel m^k)$ for all \( k \) in $ \{1, 2, \ldots, $N-1$\}$.

\subsection{Initialisation}
\label{cluster_initialization}
In the training phase of DC algorithms, the quality of the initial centroids can directly affect the quality of the final clusters. Poor initialization can lead to poorly formed clusters, such as unevenly sized, too small, or too large. 
Existing DC methods \cite{DBLP:conf/www/Bo0SZL020, DBLP:conf/aaai/LiuTZLSYZ22, DBLP:conf/aaai/TuZL0CZC21} use K-means to initialize cluster centers $\textit{c}$. However, in data integration problems dealing with highly dense and overlapping embedding representation, K-means may not be a suitable initialization (see ablation study in Section \ref{app:ablation_study}) for the following reasons: (i) In densely packed and overlapping data dimensions, the Euclidean distance between $\textit{c}$ and $\textit{z}$ becomes uniform, which leads to poor cluster initialization. 
(ii) In a dense data space, many sub-optimal solutions exist, and K-means is not optimal enough to avoid local minima. (iii) K-means initialization is based on hard assignment, where the centers are randomly initialized. However, dense data with overlapping clusters requires soft assignments based on probabilities. K-means initialization is often considered suitable in existing DC proposals due to the nature of the data, which is mainly sparse or semi-dense, particularly in image representations. The Euclidean distance between different points becomes more distinct in a sparse data space, resulting in well-separated clusters. This distinct separation enhances the effectiveness of K-means initialization in these contexts, facilitating the clustering process by providing a clear starting point for the algorithm.

In TableDC, we integrate the Birch algorithm ~\cite{DBLP:conf/sigmod/ZhangRL96} to initialize the clusters. Birch is a traditional hierarchical clustering algorithm that handles large and noisy databases. Birch uses a Clustering Feature tree (CF-tree) to process each data point. A CF-tree is a set of nodes, and each node consists of a triplet that efficiently stores information on data points in a compressed format to the cluster instead of processing a complete dataset. The triplet includes the number of data points per cluster, squared, and linear sum. In a dense space, each node of the CF-tree represents aggregated properties of each data point instead of the distance between data points, which enables Birch initialization to avoid close proximity and high overlaps. Since Birch is hierarchical, the multiple levels of the CF-tree can effectively capture the different granularities of clusters even if the overlaps exist on one level. 
We compare different cluster center initialization techniques in the ablation study (see Section \ref{app:cluster_initialization}). The pseudo-code for the initialization of $\textit{c}$ using Birch is given in Algorithm \ref{algo2}).

\begin{algorithm}[t]
\caption{Pseudo-code for initialization of $\textit{c}$ using Birch}
\label{algo2}
\begin{algorithmic}[1] 
\REQUIRE $\textit{z}$: data points; $\textit{K}$: number of clusters; $\textit{\textit{T}}$: threshold; $\textit{B}$: branching factor (the maximum number of child nodes); $\textit{L}$: The maximum number of entries in a leaf node
\ENSURE $\textit{c}$: cluster centers  and cluster assignments for each $\textit{z}$
\STATE  Create an empty CF-tree with $\textit{T}$, $\textit{B}$, and $\textit{L}$.
\STATE Initialize Birch clustering with $\textit{K}$
\FOR{each $\textit{z}$ in the dataset}
        \STATE Traverse the CF-tree to find a suitable corresponding leaf node
        \STATE  Update the CF-tree in the path from the root to the last leaf node at each level
        \IF{the value of leaf node $>$ $\textit{\textit{T}}$}
        \STATE Split the leaf node
        \ENDIF
        \ENDFOR
        \STATE Compute clusters $\textit{k}$ and assign a cluster index to each $\textit{z}$
\FOR{each $\textit{k}$}        
        \STATE Compute $\textit{c}$ as a mean of $\textit{z}$ assigned to $\textit{k}$
        \STATE Add $\textit{c}$ to the list of cluster centers     
\ENDFOR
\RETURN  $c_i = c$ 
\end{algorithmic}
\end{algorithm}

\section{Evaluation}
This section evaluates TableDC using existing benchmarks with a variety of different embedding approaches, in comparison with other clustering algorithms and bespoke solutions. 


\subsection{Experimental setup}

\subsubsection{Datasets}
\label{datasets}
TableDC is evaluated on six benchmark datasets widely adopted for data integration problems, specifically schema inference, domain discovery and entity resolution.  A brief description and statistical detail of each data set are given below and in Table \ref{tab:datasets:1}.

\begin{itemize}
    \item T2D Entity-Level Gold standard (T2D)~\cite{DBLP:conf/edbt/RitzeB17} is a widely used collection of web tables with mappings to their corresponding DBpedia class. Regarding schema inference, T2D can be used to align the schema of web tables (structure and format) with the schema used in DBpedia. In T2D, we excluded all tables categorized under the DBpedia class 'Thing'. This is because the 'Thing' category is mapped to over 50\% of the web tables, leading to a significant risk of data imbalance.
    \item Table Union Search (TUS) benchmark \cite{DBLP:journals/pvldb/NargesianZPM18}, refers to finding unionable tables for a given query table. However, for clustering application, the problem can be redefined as each cluster shares unionable tables. 
    \item MusicBrainz \cite{DBLP:conf/adbis/SaeediPR17} dataset is derived from authentic records of songs sourced from the MusicBrainz database. It combines the data from five distinct sources, where approximately 50\% of the original song records are duplicated across two to five sources.
    \item Geographic Settlements (GeoSet) \cite{DBLP:conf/adbis/SaeediPR17} represents a collection of geographical real-world entities from four different sources (Geonames, Freebase, DBpedia, and NYTimes).
    \item Di2KG (Camera and Monitor)\footnote{\url{http://di2kg.inf.uniroma3.it/datasets.html}} contains a set of cameras and monitors web records scraped from multiple e-commerce web pages. Both datasets are highly heterogeneous. For example, specification  \textit{lcd display} from \textit{www.price-hunt.com} and \textit{monitor} from \textit{www.cambuy.com.au} semantically represent the same domain despite having different syntactic structures.
\end{itemize}

We use T2D and TUS for schema inference, MusicBrainz and GeoSet for entity resolution, and Di2KG (Camera and Monitor) datasets for domain discovery.

\begin{table}[t]
  \caption{Statistics of benchmark datasets used to evaluate TableDC  for schema inference, entity resolution and domain discovery.}
  \label{tab:datasets:1}
  \begin{tabular}{ccccccccccccccccc}
	\toprule  
\begin{tabular}{cccc} 
 & \textbf{Dataset} & \textbf{Instances} & \textbf{Clusters} \\ 
 \midrule
\textbf{Tables} & web tables & 429 & 26 \\ 
 & TUS & 4248 & 37 \\ 
\textbf{Rows} & Music Brainz & 2002 & 684 \\ 
 & GeoSet & 3021 & 786 \\ 
\textbf{Columns} & Camera & 19036 & 56 \\ 
 & Monitor & 34481 & 81 \\ 
 \bottomrule
\end{tabular}
 \end{tabular}
 \end{table}

\subsubsection{Benchmark Methods}
\label{benchmar_methods}
We compared TableDC\footnote{https://github.com/hafizrauf/TableDC} with three SC and five DC clustering methods.  
The selected SC methods represent different paradigms. Similarly, the chosen DC methods represent different approaches to representation learning (self-supervised and subspace). A short description of each benchmark method is given below:

\begin{itemize}
    \item K-means ~\cite{Hartigan_1979} is a classical clustering algorithm; it assigns instances to the nearest centroid to create clusters. Centroids are updated based on the assigned points.

    \item DBSCAN \cite{DBLP:conf/kdd/EsterKSX96} identifies distinct clusters based on the data density, considering the number of neighboring points within a specified radius. DBSCAN is particularly good with data having variation in densities.

    \item Birch ~\cite{DBLP:conf/sigmod/ZhangRL96} is a hierarchical clustering algorithm built on a CF-tree to summarize the information (on each tree node) needed to cluster instances.

    \item SHGP \cite{DBLP:conf/nips/0002GW0XLH22} is a self-supervised method that shares an attention-aggregation mechanism between two modules, \textit{Att-LPA } to produce pseudo-labels and \textit{Att-HGNN}, to learn object embeddings. Both modules effectively improve each other to optimize embeddings.

    \item DCRN \cite{DBLP:conf/aaai/LiuTZLSYZ22} learns representations through AE and GCN by reducing the information correlation to improve the discriminative property of the resulting features.

    \item DFCN \cite{DBLP:conf/aaai/TuZL0CZC21} focuses on the fusion of representation learning of AE and graph neural networks. It works on a dynamic fusion mechanism to refine the graph neural network structure to better integrate with AE representation learning.

    \item EDESC \cite{DBLP:conf/cvpr/CaiFGWZ022} EDESC  is a subspace representation-based learning method that iteratively refines the bases of the subspace derived from deep representations. 

    \item SDCN \cite{DBLP:conf/www/Bo0SZL020} effectively utilizes the structural information in the AE representation learning. The dual self-supervised mechanism effectively updates the model parameters to produce clustering-specific representations.
\end{itemize}

\begin{table*}[h]
  \caption{Comparing schema inference clustering results (TableDC vs. existing DC methods). The best results are shown in bold. Representations marked with $*$ are obtained on instance-level data and $TT^*$ refers to TabTransformer.}
  \label{tab:siresult:1}
  \begin{tabular}{ccccccccccccccccc}
	\toprule  
Method & \multicolumn{8}{c}{TUS} & \multicolumn{8}{c}{Web tables} \\ 
 & \multicolumn{2}{c}{SBERT} & \multicolumn{2}{c}{FastText} & \multicolumn{2}{c}{$TT^*$} & \multicolumn{2}{c}{$SBERT^*$} & \multicolumn{2}{c}{SBERT} & \multicolumn{2}{c}{USE} & \multicolumn{2}{c}{$TT^*$} & \multicolumn{2}{c}{$SBERT^*$} \\
 \midrule  
 & ARI & ACC & ARI & ACC & ARI & ACC & ARI & ACC & ARI & ACC & ARI & ACC & ARI & ACC & ARI & ACC \\ 
K-means & 0.73 & 0.79 & 0.63 & 0.69 & 0.22 & 0.35 & 0.46 & 0.56 & 0.27 & 0.45 & 0.19 & 0.40 & -0.013 & 0.27 & 0.32 & 0.51 \\ 
DBSCAN & 0.17 & 0.47 & 0.01 & 0.25 & 0.02 & 0.26 & 0.09 & 0.38 & 0.0 & 0.29 & 0.0 & 0.29 & -0.007 & 0.26 & 0.0 & 0.29 \\ 
Birch & 0.22 & 0.40 & 0.0 & 0.20 & 0.18 & 0.35 & 0.54 & 0.58 & 0.33 & 0.49 & 0.22 & 0.40 & -0.013 & 0.27 & 0.46 & 0.61 \\ 
SHGP & 0.61 & 0.72 & 0.43 & 0.56 & 0.05 & 0.19 & 0.13 & 0.29 & 0.08 & 0.29 & 0.07 & 0.28 & -0.0031 & 0.27 & 0.09 & 0.30 \\ 
DCRN & 0.50 & 0.59 & $\infty $ & $\infty $ & 0.018 & 0.24 & 0.15 & 0.26 & 0.07 & 0.25 & 0.03 & 0.25 & -0.02 & 0.25 & 0.17 & 0.36 \\ 
DFCN & 0.55 & 0.65 & 0.43 & 0.52 & 0.20 & 0.37 & 0.34 & 0.43 & 0.24 & 0.40 & 0.19 & 0.33 & -0.0006 & 0.24 & 0.37 & 0.48 \\ 
EDESC & 0.28 & 0.35 & 0.22 & 0.32 & 0.14 & 0.30 & 0.36 & 0.43 & 0.26 & 0.44 & 0.22 & 0.37 & \textbf{0.08} & 0.25 & -0.04 & 0.26 \\ 
SDCN & 0.60 & 0.57 & 0.53 & 0.47 & \textbf{0.25} & \textbf{0.40} & 0.45 & 0.49 & 0.20 & 0.34 & 0.08 & 0.33 & 0.0 & \textbf{0.29} & 0.30 & 0.42 \\ 
TableDC & \textbf{0.88} & \textbf{0.87} & \textbf{0.73} & \textbf{0.76} & 0.24 & 0.37 & \textbf{0.63} & \textbf{0.66} & \textbf{0.62} & \textbf{0.65} & \textbf{0.25} & \textbf{0.42} & -0.006 & 0.26 & \textbf{0.61} & \textbf{0.65} \\ 
\bottomrule 
\end{tabular}
\normalsize
\end{table*}

\subsubsection{Representations}
Each data integration task has an associated embedding model for representing tables, rows or columns which is integrated into the DC method for clustering. We use six different embedding models, each of which is specific to one or more data integration tasks; details are given below:

  \textbf{Schema Inference}: We use pre-trained embeddings to get the representations of raw data. We categorized these into two categories. When we consider headers, we use SBERT ~\cite{Reimers_2019}, FastText ~\cite{DBLP:conf/lrec/GraveBGJM18} and USE \cite{DBLP:conf/emnlp/CerYKHLJCGYTSK18}; with instances, we use TabTransformer and SBERT. SBERT uses siamese and triplet network structures to extract semantically meaningful embeddings from sentences. FasText is specialized in extracting morphological information and considers subwords, representing each word as a bag of character n-grams. Like SBERT, USE is also a sentence encoder specializing in capturing the semantic meanings of sentences. Considering the instances along with the header is a more tricky task and requires more robust transformers. We fine-tuned TabTransformer~\cite{DBLP:journals/corr/abs-2012-06678} on both datasets to benefit the categorical and contentious features in the dataset. We also encoded instance data with SBERT; we convert each row of the table into a sentence sequence, where each row is represented as a sequence of its cell values appended with \textit{[SEP]} token for SBERT to separate segments and maintain the row boundaries in the text. \textit{[SEP]} carries the table structure within the BERT model in the encoded form.

 \textbf{Entity Resolution}: We train EmbDi~\cite{DBLP:conf/sigmod/CappuzzoPT20} on both datasets to embed rows and used pre-trained SBERT. EmbDi is a graph-based representation method based on tripartite connectivity within a graph of table elements to extract structural information. A tripartite connection contains a column node (attribute representation), a value node (unique value representation), and a row node (unique tuple token).

 \textbf{Domain Discovery}: Similar to schema inference, we embed columns in two categories, with and without column headers. We used SBERT and T5 \cite{DBLP:journals/jmlr/RaffelSRLNMZLL20} pre-trained embeddings for both purposes. When considering instances, we embed column values using SBERT and T5 and aggregate the individual embeddings to get the final column embeddings. T5 is a widely used language model developed by Google for information retrieval tasks and generating embeddings, specifically capturing semantic information within data.

\subsection{Metrics}
We adopted two widely used standard clustering evaluation metrics, Accuracy (ACC) ~\cite{DBLP:journals/tip/YangXNYZ10} and Adjusted Rand Score (ARI) ~\cite{DBLP:conf/iccv/WuLWQLLZ19} to test the performance of TableDC and existing benchmark methods. ARI ranges from -1 to 1, where 1 indicates perfect matching, 0 indicates random labeling and negative values indicate worse than random labeling. ACC ranges from 0 to 1.

\subsection{Parameter setting}

TableDC consists of four AE layers and is trained using the Adam optimizer. The loss function combines KL-divergence (to minimize the Mahalanobis distance in self-supervised learning) and mean squared error loss for reconstruction. The latent space size is fixed at 100. In order to be consistent with the existing DC benchmarks, we pre-trained TableDC with AE on 30 epochs for schema inference and domain discovery. Pretraining provides a good starting point for the network parameters, avoiding random initialization and local minima during the clustering phase, which leads to faster convergence. Since entity resolution involves many clusters, the CF-tree in Birch initialization needs more nodes with detailed cluster summaries, each representing a smaller and more defined cluster. We have pre-trained TableDC on 100 epochs for entity resolution. Training epochs for schema inference are fixed to 200, domain discovery to 100, and entity resolution to 50. We run all existing benchmark methods on the same number of epochs adopted for TableDC, excluding the internal parameters for which we use the originally published values for each existing method, such as the number and size of each layer.  We tuned the threshold $\textit{\textit{T}}$ (the maximum range of radius of a cluster in the CF-tree) when initiating the centroids in TableDC for those experiments where the number of subclusters found on each CF-tree node is less to be able to summarize cluster information. We adopted a grid search on threshold $\textit{\textit{T}}$ and divided the number of sub-clusters unless the CF-tree becomes stable. Some existing methods need K-means for centroids initialization and clustering assignments. To avoid extreme cases, we initialize 20 times and choose the best solution.

\subsection{Schema inference clustering results}
Schema inference as a clustering problem can be defined as the clustering of tables that share a similar schema. Table \ref{tab:siresult:1} shows the clustering results of schema inference on Table Union Search (TUS) and web tables datasets encoded with several representations. 

We observe the following:
(i) \textit{TableDC obtained the best results over both datasets when considering schema-level data} (SBERT, FastText, USE) compared to all benchmark methods, including SC. TableDC improves the clustering performance (with SBERT) by 0.28 and 0.27 ARI and 0.30 and 0.15 ACC, compared to SDCN and SHGP, respectively.
(ii) \textit{TableDC exhibits good but not always best performance on instance-level data} (TabTransformer, SBERT) and was outperformed by SDCN by a small margin (0.01 ARI and 0.03 ACC) on the TUS dataset and by EDESC on the web tables dataset using TabTransformer. However, TableDC outperformed the other methods with SBERT, improving clustering performance by 0.31 average ARI and 0.23 average ACC on the TUS dataset.
(iii) \textit{TableDC handles overlapping clusters due to its heavy-tailed Cauchy distribution as a similarity kernel}. For example, TableDC assigns two tables that share similar schemas, \textit{RadioStation.(Radio, Country, Language)} and \textit{Country.(Country, Language, Broadcasters)} with a 0.79 cosine similarity score in the representation space to two different clusters, compared to SDCN and DFCN, which failed to handle high overlap and put them in one cluster, leading to misclassification.
(iv) \textit{TableDC provides effective distance measures in a dense space through the Mahalnobis distance measure when calculating the distance between data points}. Two instances from different GT clusters are closely packed in the latent space when the distance measured is Euclidean compared to the Mahalbonis distance, which helps the model by placing both instances apart. For example, the Mahalanobis distance between different attributes of two tables \textit{Book.(rank, title, genre, creator, date, freq)} and \textit{Film.(rank, title, year, director, overall rank)}, is 0.99 compared to the Euclidean distance, which is 0.71, indicating that during self-supervision in TableDC, both instances are further apart despite having highly overlapping values. Experimentally, TableDC correctly placed the two instances in different clusters whereas EDESC and DCRN misclassified them into one cluster due to a low Euclidean distance score. (v) \textit{TableDC focuses on semantic similarity more than distributional frequency}. For example, when considering instances on TUS with SBERT,  attributes \textit{(Day, Month)} with different frequencies (i.e., the frequency of each day) present in a column should be in the same clusters, ignoring the synthetic properties or the count of a particular day repeated in a  column. However, SDCN and DCRN consider the syntactic similarity and the frequency \textit{(Tuesday and Wednesday are the highest counts in both tables)} and incorrectly place the two tables in different clusters, in contrast with TableDC.

\begin{table}[!t]
  \caption{Comparing entity resolution clustering results (TableDC vs. existing DC methods). The best results are shown in bold.}
  \label{tab:erresult:1}
  \small
  \begin{tabular}{ccccccccc} 
  \toprule
 & \multicolumn{4}{c}{\textbf{Music Brainz}} & \multicolumn{4}{c}{\textbf{GeoSet}} \\ 
 \midrule
Method & \multicolumn{2}{c}{SBERT} & \multicolumn{2}{c}{EmbDi} & \multicolumn{2}{c}{SBERT} & \multicolumn{2}{c}{EmbDi} \\ 
 & ARI & ACC & ARI & ACC & ARI & ACC & ARI & ACC \\ 
K-means & 0.40 & 0.68 & 0.39 & 0.64 & 0.57 & 0.83 & 0.56 & 0.71 \\ 
DBSCAN & 0.0 & 0.002 & N/A & N/A & 0.0 & 0.001 & 0.0 & 0.001 \\ 
Birch & 0.56 & 0.76 & 0.41 & 0.67 & 0.56 & 0.75 & 0.59 & 0.71 \\ 
SHGP & 0.15 & 0.47 & 0.20 & 0.51 & 0.52 & 0.77 & 0.43 & 0.63 \\ 
DFCN & 0.03 & 0.31 & 0.03 & 0.33 & 0.34 & 0.59 & 0.37 & 0.57 \\ 
EDESC & -0.001 & 0.03 & 0.03 & 0.29 & 0.30 & 0.73 & 0.25 & 0.52 \\ 
SDCN & 0.002 & 0.13 & 0.06 & 0.47 & 0.05 & 0.66 & 0.52 & 0.66 \\ 
\textbf{TableDC} & \textbf{0.80} & \textbf{0.88} & \textbf{0.51} & \textbf{0.71} & \textbf{0.65} & \textbf{0.86} & \textbf{0.60} & 0.71 \\ 
\bottomrule
\end{tabular}
\normalsize
\end{table}

\begin{table*}[!h]
  \caption{Comparing domain discovery clustering results (TableDC vs. existing DC methods). The best results are shown in bold.}
  \label{tab:ddresult:1}
  \begin{tabular}{ccccccccccccc} 
  \toprule
Method & \multicolumn{6}{c}{\textbf{Camera}} & \multicolumn{6}{c}{\textbf{Monitor}} \\ 
 & \multicolumn{2}{c}{SBERT} & \multicolumn{2}{c}{$SBERT^*$} & \multicolumn{2}{c}{$T5^*$} & \multicolumn{2}{c}{SBERT} & \multicolumn{2}{c}{$SBERT^*$} & \multicolumn{2}{c}{$T5^*$} \\ 
 \midrule
 & ARI & ACC & ARI & ACC & ARI & ACC & ARI & ACC & ARI & ACC & ARI & ACC \\ 
K-means & 0.74 & 0.70 & 0.49 & 0.56 & 0.49 & 0.56 & 0.59 & 0.58 & 0.54 & 0.53 & 0.52 & 0.53 \\ 
DBSCAN & 0.73 & 0.69 & -0.005 & 0.25 & -0.005 & 0.25 & 0.27 & 0.50 & 0.007 & 0.22 & 0.006 & 0.22 \\ 
Birch & 0.76 & 0.70 & 0.78 & 0.74 & 0.69 & 0.68 & 0.55 & 0.55 & 0.60 & 0.61 & 0.53 & 0.53 \\ 
SHGP & 0.65 & 0.65 & 0.47 & 0.56 & 0.41 & 0.51 & 0.59 & 0.60 & 0.48 & 0.52 & 0.46 & 0.49 \\ 
DCRN & 0.60 & 0.60 & 0.41 & 0.44 & N/A & N/A &N/A  & N/A & N/A & N/A & N/A & N/A \\ 
DFCN & 0.77 & 0.72 & 0.64 & 0.68 & 0.57 & 0.61 & 0.59 & 0.59 & 0.53 & 0.54 & 0.50 & 0.50 \\ 
EDESC & 0.41 & 0.48 & 0.57 & 0.59 & 0.46 & 0.54 & 0.32 & 0.42 & 0.48 & 0.50 & 0.46 & 0.49 \\ 
SDCN & 0.68 & 0.67 & 0.68 & 0.67 & 0.66 & 0.63 & 0.47 & 0.52 & 0.52 & 0.54 & 0.55 & 0.54 \\ 
\textbf{TableDC} & \textbf{0.80} & \textbf{0.72} & \textbf{0.82} & \textbf{0.75} & \textbf{0.77} & \textbf{0.74} & \textbf{0.64} & \textbf{0.65} & \textbf{0.63} & \textbf{0.61} & \textbf{0.58} & \textbf{0.57} \\ 
\bottomrule
\end{tabular}
\normalsize
\end{table*}

\subsection{Entity resolution clustering results}
Entity resolution as a clustering problem can be defined as clustering multiple records that refer to the same real-world entity. Table \ref{tab:erresult:1} shows the clustering results of entity resolution on GeoSet and Music Brainz datasets encoded with several representations. 

We observe the following: (i) \textit{The best results are with TableDC on both datasets}, outperforming all other methods by 0.63 average ARI and 0.53 average ACC on Music Brainz and 0.31 average ARI and 0.24 average ACC on GeoSet using SBERT. 
(ii) \textit{TableDC performed well with SBERT compared to EmbDi,} delivering good semantic similarity performance among different records of the same real-world entity. TableDC is better with SBERT than EmbDi by 0.20 ARI and 0.17 ACC with Music Brainz, while 0.05 ARI and 0.15 ACC with GeoSet.
(iii) \textit{TableDC more effectively adopted the semantic mapping by SBERT than SHGP and DFCN by clustering contextually similar records}. For example, two records of geographical locations \textit{(name: Manchester (England), longitude: -2.23743, latitude: 53.4809)} from \textit{data.nytimes.com} and \textit{(name: manchester united kingdom, longitude: -2.24, latitude: 53.48)} from \textit{rdf.freebase.com} have a low syntactic similarity (pairwise Euclidean distance) of 0.47 and high cosine similarity (pairwise cosine distance) of 0.86. SHGP and DFCN misclassified both instances and grouped them into different clusters compared to TableDC, which forms one cluster with contextually similar records.
(iv) \textit{TableDC with EmbDi on GeoSet created fewer unary clusters (50) than SHGP (101) and EDESC (90)}, indicating that TableDC effectively balances cluster purity and cluster size. A higher ARI score with fewer unary clusters implies that TableDC is better at avoiding unnecessary fragmentation of the rows into too many small clusters. (v) \textit{Like schema inference, in entity resolution, existing clustering methods failed to form clusters based on the context of a particular text}. For example, two different songs with some similar values of attributes are clustered together by SDCN and SHGP with EmbDi, i.e. \textit{(title: Jabberwock,  length: 292706, year: 2010,  language: French)} and \textit{(title: Bubble Star,  length: 244346, year: 2004,  language: French)}. Both records share the same \textit{language} and numeric values with a similar range encoded by EmbDi despite having different \textit{titles}.  TableDC managed to put them in different clusters based on the context of both records despite them having several similar attribute values.

\subsection{Domain discovery clustering results}

Domain discovery as a clustering problem can be defined as a clustering of the columns from a collection of datasets that refer to the same domain. Table \ref{tab:ddresult:1} shows the clustering results of domain discovery on Camera and Monitor datasets encoded with several representations.

We observe the following: 
(i) \textit{TableDC obtained the best results with SBERT and T5 when considering column values of the Camera dataset} outperforming other methods by an average of 0.32 and 0.18 with SBERT and 0.30 and 0.20 with T5 on ACC and ARI, respectively.
(ii) \textit{TableDC also led on the Monitor dataset} by an average of 0.18 and 0.11 with SBERT and an average of 0.14 and 0.09 with T5 on ARI and ACC, respectively.
(iii) \textit{TableDC shows a superior capacity for learning and interpreting the column's context in the latent space compared to other methods}. TableDC's ability to learn and utilize complex patterns shows the better integration of the embeddings obtained using T5 on the Camera dataset. For example, the cosine similarity of the T5 vectors of two columns \textit{(image sensor: CMOS)} and \textit{(optical sensor: CMOS)} is 0.61, and TableDC learns the context and assigns both columns to the same cluster, compared to SDCN and DFCN, which incorrectly assigns them to different clusters.
(iv) \textit{The clustering performance of existing methods is affected by the length of columns compared to TableDC, which handles the columns uniformly regardless of the length}. For example, two columns of the Camera dataset with different lengths  \textit{(camera color: silver gray/ black/ red/ silver)} and \textit{(color: black)} are categorized in different clusters by TableDC and DCRN with SBERT; however, TableDC overrides the column structure and correctly puts them together in the same cluster. In the experimental evaluation, some experiments with DCRN have not managed to scale on available resources and are reported as \textit{N/A}.



\subsection{Comparison with bespoke solutions}
\label{sec:bespoke}

Previous experiments have compared TableDC with other SC and DC algorithms, using the same representations. This section complements such results by comparing TableDC with existing state-of-the-art solutions.

\subsubsection{Schema Inference}

Although there are many proposals for schema inference techniques, as surveyed in~\cite{DBLP:journals/vldb/CebiricGKKMTZ19,Kellou-Menouer-22}, few of these are for tabular data. Thus the comparators in this section use state-of-the-art techniques for table similarity along with standard clustering algorithms.  Specifically for table similarity, we use:
\begin{itemize}
    \item$D^3L$~\cite{DBLP:conf/icde/BogatuFP020}, which combines the results of several largely syntactic methods for comparing column headers and values into an overall table similarity score. $D^3L$ is used with the K-means for clustering because it exhibits the best results compared to Birch and DBSCAN.
    \item Starmie ~\cite{DBLP:journals/pvldb/FanWLZM23}, which uses self-supervised contrastive learning to fine tune ROBERTa ~\cite{DBLP:journals/corr/abs-1907-11692} language models for column similarity; table similarity is then supported by combining these column similarities. Starmie is used with the  connected component algorithm ~\cite{DBLP:journals/pvldb/FanWLZM23} as in the original paper.
\end{itemize}

\subsubsection{Entity Resolution}

There are many proposals for entity resolution techniques, as surveyed in~\cite{DBLP:journals/csur/ChristophidesEP21,DBLP:journals/tkde/ElmagarmidIV07}, though many of these consider pairwise similarity rather than producing clusters. Thus we specifically select a framework that can produce clusters of instances, namely JedAI ~\cite{DBLP:journals/is/PapadakisMGSTGB20,DBLP:journals/is/Mandilaras0GSTG21}, which provides several workflows for entity resolution. Within JedAI, we use the schema agnostic workflow, as TableDC is also schema agnostic; in essence, both techniques can be used to look for duplicates in tables with different structures.  JedAI workflows can also be configured to support different comparison metrics, and results are reported for Jaccard, Cosine and Dice.

\subsubsection{Domain Discovery}

There are relatively few fully unsupervised proposals for Domain Discovery.  Here we use one bespoke method, namely D4~\cite{DBLP:journals/pvldb/OtaMFS20}, which identifies domains by considering overlaps between column instances, and as in Schema Inference we use Starmie~\cite{DBLP:journals/pvldb/FanWLZM23} along with an SC algorithm.  The comparison with Starmie involves comparing a language model that has been fine-tuned for similarity with a language model in TableDC that is fine-tuned for clustering.

\subsubsection{Observations} 
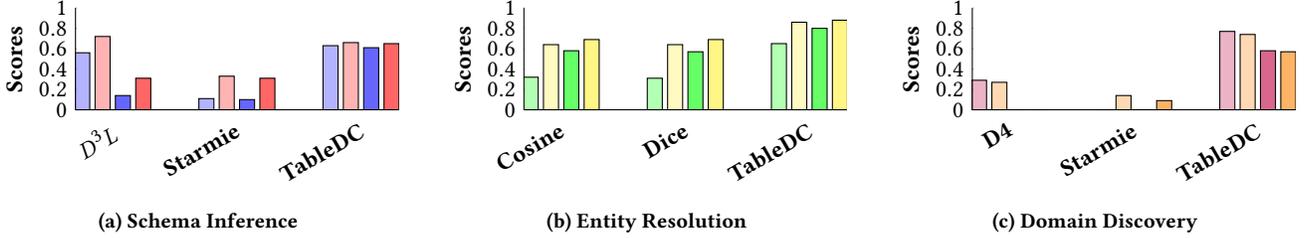
\begin{figure*}[!htbp]
\centering

\begin{tikzpicture}
    \matrix[font=\bfseries, row sep=0.5em, column sep=1em] at (0,0) {
        \node[fill=blue!30, label=right:ARI (TUS)] {}; &
        \node[fill=red!30, label=right:ACC (TUS)] {}; &
        \node[fill=blue!60, label=right:ARI (web tables)] {}; &
        \node[fill=red!60, label=right:ACC (web tables)] {}; &
        \node[fill=green!30, label=right:ARI (GeoSet)] {}; &
        \node[fill=yellow!30, label=right:ACC (GeoSet)] {}; \\
        \node[fill=green!60, label=right:ARI (Music Brainz)] {}; &
        \node[fill=yellow!60, label=right:ACC (Music Brainz)] {}; &
        \node[fill=purple!30, label=right:ARI (Camera)] {}; &
        \node[fill=orange!30, label=right:ACC (Camera)] {}; &
        \node[fill=purple!60, label=right:ARI (Monitor)] {}; &
        \node[fill=orange!60, label=right:ACC (Monitor)] {}; \\
    };
\end{tikzpicture}

\vspace{0.5cm}

\begin{subfigure}{0.33\textwidth}
    \centering
    \begin{tikzpicture}
        \begin{axis}[
            ybar,
            bar width=.2cm,
            width=\textwidth,
            height=0.5\textwidth,
            enlarge x limits=0.15,
            ylabel={\textbf{Scores}},
            symbolic x coords={$D^3L$, Starmie, TableDC},
            xtick=data,
            xticklabel style={rotate=30, anchor=north east, font=\bfseries},
            ymin=0, ymax=1,
            axis lines*=left,
            axis on top,
            tickwidth=0pt,
            ylabel near ticks,
            xlabel near ticks,
            yticklabel style={/pgf/number format/fixed, font=\bfseries},
        ]
        \addplot[fill=blue!30] table[x=Method11, y=ARI1] {\datasetoneandtwo};
        \addplot[fill=red!30] table[x=Method11, y=ACC1] {\datasetoneandtwo};
        \addplot[fill=blue!60] table[x=Method11, y=ARI2] {\datasetoneandtwo};
        \addplot[fill=red!60] table[x=Method11, y=ACC2] {\datasetoneandtwo};
        \end{axis}
    \end{tikzpicture}
    \caption{\textbf{Schema Inference}}
\end{subfigure}
\hfill
\begin{subfigure}{0.33\textwidth}
    \centering
    \begin{tikzpicture}
        \begin{axis}[
            ybar,
            bar width=.2cm,
            width=\textwidth,
            height=0.5\textwidth,
            enlarge x limits=0.15,
            ylabel={\textbf{Scores}},
            symbolic x coords={Cosine, Dice, TableDC},
            xtick=data,
            xticklabel style={rotate=30, anchor=north east, font=\bfseries},
            ymin=0, ymax=1,
            axis lines*=left,
            axis on top,
            tickwidth=0pt,
            ylabel near ticks,
            xlabel near ticks,
            yticklabel style={/pgf/number format/fixed, font=\bfseries},
        ]
        \addplot[fill=green!30] table[x=Method1111, y=ARI1] {\datasetthreeandfour};
        \addplot[fill=yellow!30] table[x=Method1111, y=ACC1] {\datasetthreeandfour};
        \addplot[fill=green!60] table[x=Method1111, y=ARI2] {\datasetthreeandfour};
        \addplot[fill=yellow!60] table[x=Method1111, y=ACC2] {\datasetthreeandfour};
        \end{axis}
    \end{tikzpicture}
    \caption{\textbf{Entity Resolution}}
\end{subfigure}
\hfill
\begin{subfigure}{0.33\textwidth}
    \centering
    \begin{tikzpicture}
        \begin{axis}[
            ybar,
            bar width=.2cm,
            width=\textwidth,
            height=0.5\textwidth,
            enlarge x limits=0.15,
            ylabel={\textbf{Scores}},
            symbolic x coords={D4, Starmie, TableDC},
            xtick=data,
            xticklabel style={rotate=30, anchor=north east, font=\bfseries},
            ymin=0, ymax=1,
            axis lines*=left,
            axis on top,
            tickwidth=0pt,
            ylabel near ticks,
            xlabel near ticks,
            yticklabel style={/pgf/number format/fixed, font=\bfseries},
        ]
        \addplot[fill=purple!30] table[x=Method222, y=ARI1] {\datasetfiveandsix};
        \addplot[fill=orange!30] table[x=Method222, y=ACC1] {\datasetfiveandsix};
        \addplot[fill=purple!60] table[x=Method222, y=ARI2] {\datasetfiveandsix};
        \addplot[fill=orange!60] table[x=Method222, y=ACC2] {\datasetfiveandsix};
        \end{axis}
    \end{tikzpicture}
    \caption{\textbf{Domain Discovery}}
\end{subfigure}

\caption{TableDC vs. bespoke solutions for each problem. TableDC is integrated with SBERT in (a) and (b) and T5 in (c). In (b), Jaccard and Cosine are different similarity metrics of the JedAI framework ~\cite{DBLP:journals/is/PapadakisMGSTGB20}.}
    \label{fig:bespoke}
\end{figure*}

We compare two proposals for each problem that use techniques published in top outlets (ICDE, PVLDB, Information Systems) since 2020. The results of the comparison of TableDC with bespoke solutions are presented in Figure \ref{fig:bespoke}, which reports the ARI and ACC for two datasets for each of \textit{schema inference}, \textit{entity resolution} and \textit{domain discovery}. We can observe that, for Schema inference, TableDC consistently outperforms $D^3L$ and Starmie for web tables. For TUS, $D^3L$ obtains 0.06 higher ACC than TableDC, while TableDC is still ahead with 0.07 ARI. This indicates that TableDC is good at identifying the overall clustering structure, but $D^3L$ is better at assigning the exact labels correctly. TableDC shows its ability to achieve good results on entity resolution compared to JedAI with different distance similarity measures. JedAI with cosine distance is second best with 0.58 ARI and 0.69 ACC on the Music Brainz dataset. Lastly, TableDC outperformed D4 (0.29 ARI and 0.27 ACC) in column clustering due to its ability to learn contextualized column embeddings effectively, providing more accurate clustering, with 0.77 ARI and 0.74 ACC on the Camera dataset. Overall, we can see that in the 12 experiments, TableDC provides the best results in 11 cases, thereby demonstrating that, though generic, TableDC compares well with recent, specialized state-of-the-art proposals.   

\pgfplotstableread[row sep=\\,col sep=&]{
    Method11 & ARI1 & ACC1 & ARI2 & ACC2 \\
    $D^3L$ & 0.56	&0.72	&0.14	&0.31 \\
    Starmie & 0.11&	0.33	&0.1	&0.31 \\
    TableDC & 0.63	&0.66	&0.61	&0.65 \\
}\datasetoneandtwo

\pgfplotstableread[row sep=\\,col sep=&]{
    Method1111 & ARI1 & ACC1 & ARI2 & ACC2 \\
    Cosine &0.32	&0.64	&0.58	&0.69 \\
    Dice & 0.31	&0.64&	0.57	&0.69 \\
    TableDC & 0.65&	0.86	&0.8	&0.88\\
}\datasetthreeandfour

\pgfplotstableread[row sep=\\,col sep=&]{
    Method222 & ARI1 & ACC1 & ARI2 & ACC2 \\
    D4 & 0.29&	0.27&	0	&0 \\
    Starmie & -0.007	&0.14	&0.001	&0.09 \\
    TableDC & 0.77	&0.74	&0.58&	0.57\\
}\datasetfiveandsix

\subsection{Scalability}

\begin{figure}[t]
  \centering
  \includegraphics[width=0.9\linewidth,keepaspectratio]{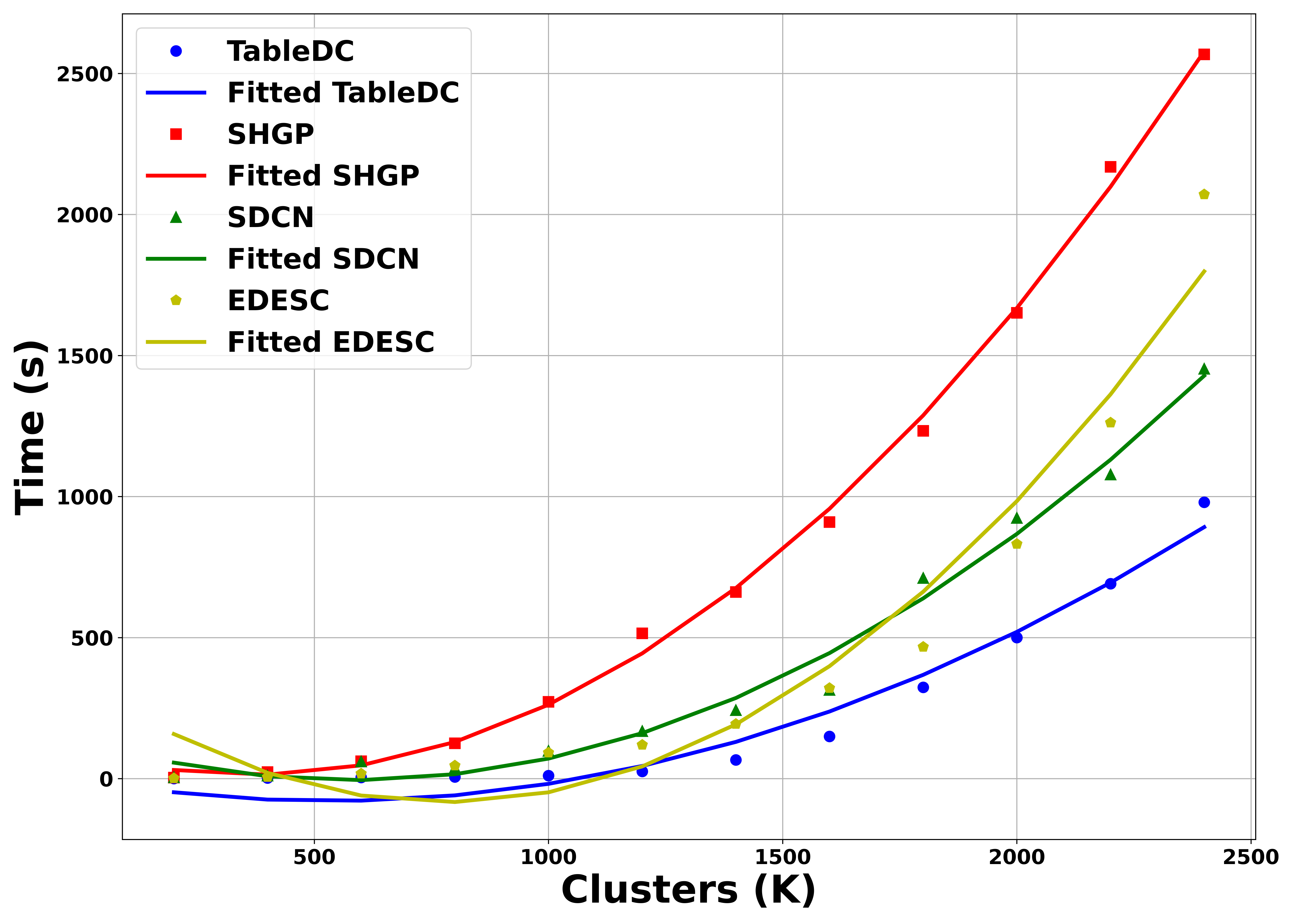}
  \caption{Scalability comparison of TableDC with existing DC approaches with respect to the number of clusters $\mathbb{K}$. DFCN and DCRN are not included in the comparison because we have not managed to run both on the available hardware resources with a high number of clusters.}
  \label{fig:scalability}
\end{figure}

We observe that existing DC proposals struggled to scale to large numbers of clusters $\mathbb{K}$ (see Figure \ref{fig:scalability}). To compare the scalability of different DC methods, we used Music Brainz, which was scaled to large numbers of clusters, up to $\mathbb{K} = 2400$. The times are reported running on Nvidia A100 GPU with 80GB GPU RAM.


Most existing DC algorithms use GCN to create an adjacency matrix to capture underlying relationships among different features. This leads to a sparse adjacency matrix in scenarios where the representation is not dense. The complexity of GCN-based representation learning, assuming a sparse adjacency matrix, is proportional to the number of vertices and edges. We often deal with dense representations with overlapping distances in data management applications. For dense representations, the multiplication between an adjacency matrix $\mathbb{A}$ (size $\mathbb{N} \times \mathbb{N}$) and a feature matrix (size $\mathbb{N} \times \mathbb{D}$) becomes more computationally intensive as $\mathbb{K}$ and $\mathbb{D}$ increase, leading to greater complexity.

In contrast, TableDC does not rely on GCN to capture relationships but instead uses Mahalanobis distance to determine correlations between different features. TableDC employs an autoencoder with a linear relationship between the number of data points $\mathbb{N}$ and the size of the hidden layer $\mathbb{H}$. The self-supervised module, which calculates the differences between cluster centroids from dense representations, maintains a manageable computational load because the calculation is around the distance from each data point to every cluster centroid. TableDC's efficiency is optimized using Cholesky decomposition to compute the Mahalanobis distance. This method efficiently handles dense representations by decomposing the covariance matrix, allowing for faster computation of distances \cite{Golub_2013}. Additionally, the integration of a Cauchy distribution kernel normalizes these distances and applies softmax operations, ensuring that the computational overhead remains low even as the number of clusters $\mathbb{K}$ increases.

\section{Ablation study}
\label{app:ablation_study}
This section presents an ablation study to investigate the impact of three components: (i) cluster initialization, (ii) loss function, and (iii) different distance measures and similarity kernels on self-supervised learning.

\subsection{Cluster Initialization}
\label{app:cluster_initialization}
\begin{figure}[t]
\centering
\begin{tikzpicture}
    \begin{axis}[
        ybar,
        bar width=.18cm,
        width=0.5\textwidth,
        height=0.30\textwidth,
        enlarge x limits=0.15,
        ylabel={\textbf{ARI}},
        symbolic x coords={K-means, K-means++, Bisecting K-Means, GMM, Birch},
        xtick=data,
        xticklabel style={rotate=30, anchor=north east,font=\bfseries},
        ymin=0, ymax=1,
        legend style={at={(0.5,0.9)}, anchor=south, legend columns=2, draw=none, font=\bfseries},
        axis lines*=left,
        axis on top,
        tickwidth=0pt,
        ylabel near ticks,
        xlabel near ticks,
        yticklabel style={/pgf/number format/fixed, font=\bfseries},
    ]
    \addplot[fill=green!30, bar width=0.2cm] coordinates {(K-means, 0.36) (K-means++, 0.28) (Bisecting K-Means, 0.31) (GMM, 0.28) (Birch, 0.62)};
     \addplot[fill=red!30, bar width=0.2cm] coordinates {(K-means, 0.02) (K-means++, 0.04) (Bisecting K-Means, 0.02) (GMM, 0.58) (Birch, 0.60)};
    \addplot[fill=blue!30, bar width=0.2cm] coordinates {(K-means, 0.63) (K-means++, 0.72) (Bisecting K-Means, 0.69) (GMM, 0.68) (Birch, 0.82)};
    
    \legend{Schema Inference,Entity Resolution,Domain Discovery}
    \end{axis}
\end{tikzpicture}
\caption{Impact of different cluster initialization methods on clustering performance. We used SBERT on web tables for schema inference, EmbDi on GeoSet for entity resolution, and SBERT on Camera for Domain Discovery.}
\label{fig:initialization}
\end{figure}
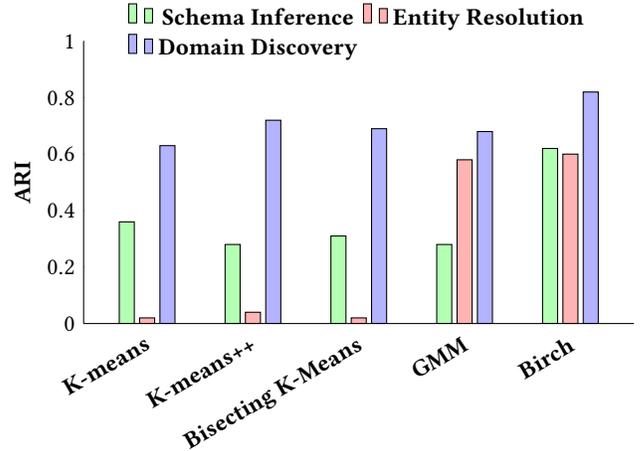

We compare several cluster centroid initialization approaches with Birch initialization to investigate how cluster initialization affects clustering performance. We record ARI using different initialization methods to evaluate the cluster quality. Figure \ref{fig:initialization} shows a bar chart comparison of TableDC trained over different initialization methods, including K-means, which has been used significantly in existing DC algorithms.

We can observe from Figure \ref{fig:initialization} that TableDC with Birch initialization provides the best results for all problems. 
The higher ARI score of TableDC with Birch initialization indicates the accurate identification of homogeneous clusters and clear separation between different clusters. K-means++ initialization appeared as the second best in domain discovery. From the comparative analysis, we also observed that with Birch initialization, the points are more widely dispersed around the centroid, indicating enhanced cluster compactness (how tightly the data points are grouped around the centroid of the cluster) and homogeneity (the similarity of data points within the same cluster). 

\begin{figure}[h]
     \centering
     \begin{subfigure}[b]{0.49\linewidth}
         \centering
         \includegraphics[width=\linewidth]{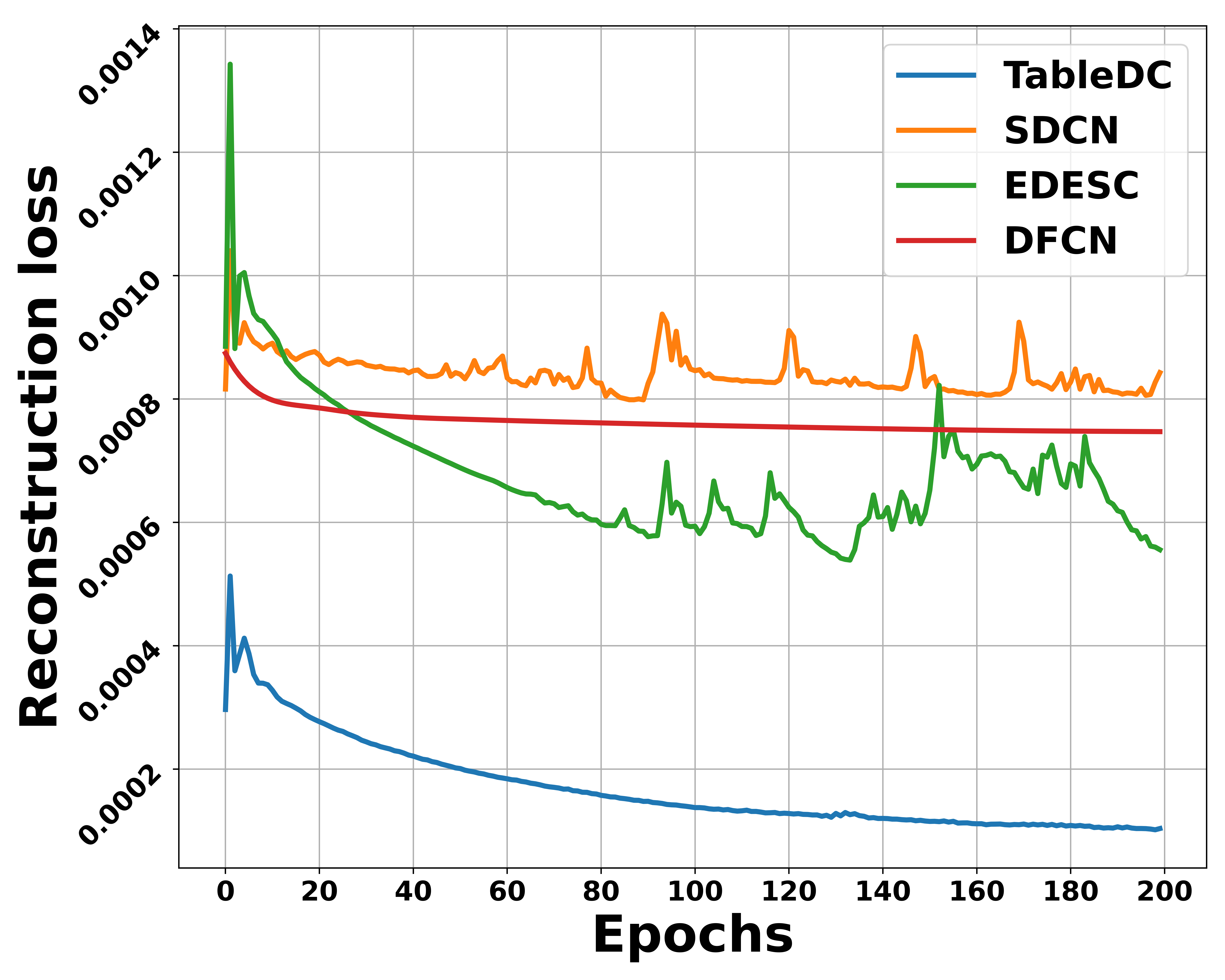}
         \caption{$re_{loss}$ minimization}
         \label{fig:loss_function1}
     \end{subfigure}
     \begin{subfigure}[b]{0.49\linewidth}
         \centering
         \includegraphics[width=\linewidth]{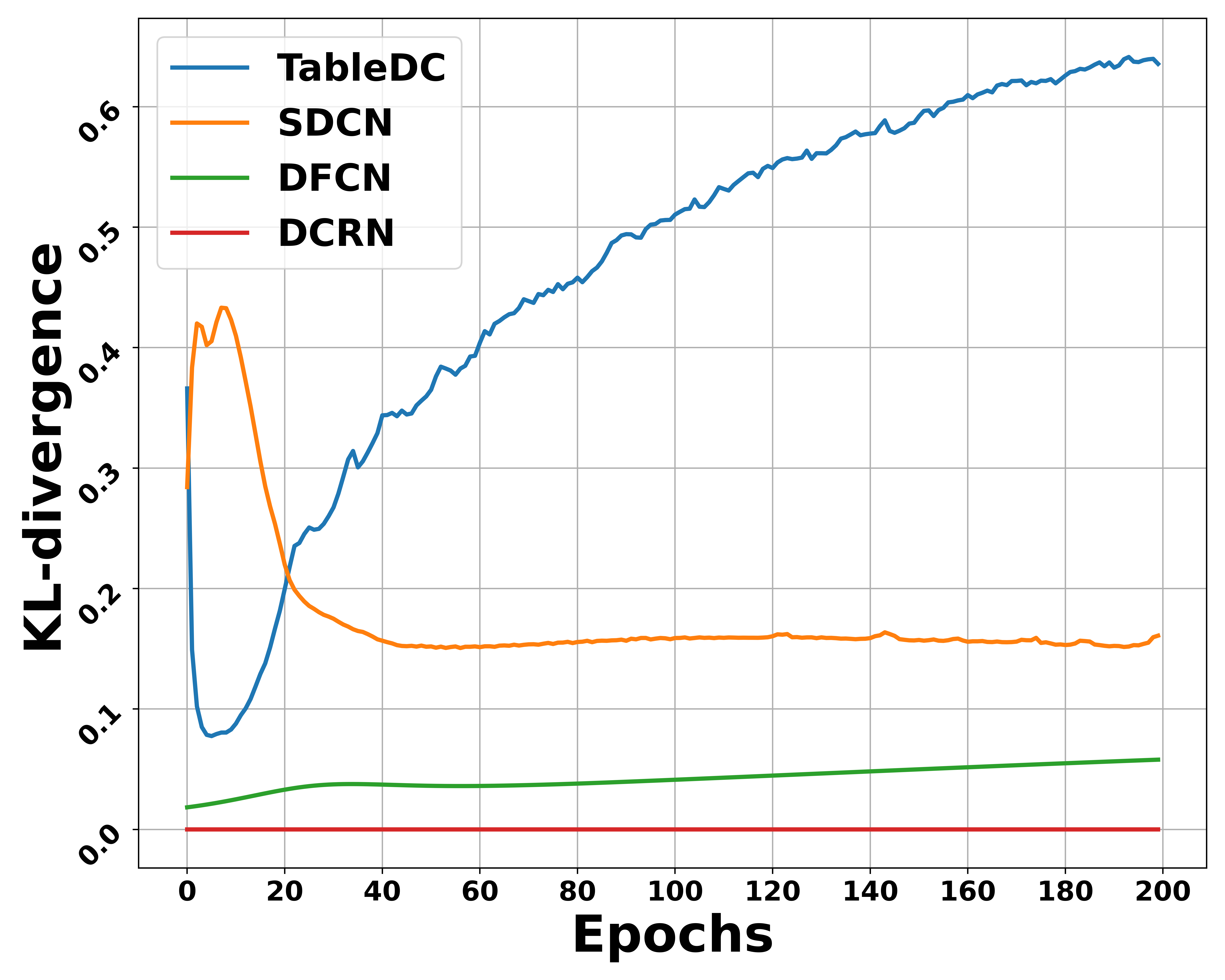}
         \caption{divergence of $q$ from $p$}
         \label{fig:loss_function2}
     \end{subfigure}
     \caption{$re_{loss}$ and $\text{KL}_\text{div}$ (between $\textit{q}$ and $\textit{p}$) comparative analysis of TableDC with benchmark methods for schema inference on web tables data. }
     \label{fig:loss_function}
\end{figure}

\begin{table*}[t]
	\caption{TableDC clustering results on different distance metrics and similarity kernels used in the self-supervised module. We used SBERT (schema only) on web tables and Monitor datasets for schema inference and domain discovery. We adopted SBERT (values only) for entity resolution on the MusicBrainz dataset.}
	\label{tab:ablation:kernel}
\begin{tabular}{cccccccccc}
	\toprule 
\textbf{} & \multicolumn{3}{c}{Schema Inference} & \multicolumn{3}{c}{Entity Resolution} & \multicolumn{3}{c}{Domain Discovery} \\ 
 \midrule
\textbf{Distance } & Euclidean & Cosine & Mahalanobis & Euclidean & Cosine & Mahalanobis & Euclidean  & Cosine & Mahalanobis \\ 
ARI & 0.49 & 0.30 & \textbf{0.62} & 0.0014 & 0.0014 & \textbf{0.88} & 0.56 & 0.30 & \textbf{0.64} \\ 
ACC & 0.62 & 0.51 & \textbf{0.65} & 0.0599 & 0.0300 & \textbf{0.87} & 0.57 & 0.41 & \textbf{0.65} \\ 
\textbf{Distribution} & Student's t & Normal & Cauchy & Student's t & Normal & Cauchy & Student's t & Normal & Cauchy \\ 
ARI & 0.52 & 0.45 & \textbf{0.62} & 0.37 & 0.0 & \textbf{0.88} & 0.54 & 0.56 & \textbf{0.64} \\ 
ACC & 0.60 & 0.64 & \textbf{0.65} & 0.69 & 0.0025 & \textbf{0.87} & 0.06 & 0.0 & \textbf{0.65} \\ 
\bottomrule
\end{tabular}
\end{table*}


\subsection{Loss Function}
\label{app:loss_function}

To empirically assess the impact of loss functions in TableDC and existing benchmarks for data management problems, we use schema inference and web table data to plot \textit{(i)} $re_{loss}$ and \textit{(ii)} $\text{KL}_\text{div}$ (see Figure \ref{fig:loss_function}). 
A well-trained AE with effective $re_{loss}$ minimization, leads to embeddings where clusters are more distinct and separable, specifically in the cases of high density and overlap in the original embeddings.  We can observe from Figure \ref{fig:loss_function} that the $re_{loss}$ of TableDC is significantly reduced and more consistent compared to the benchmark methods, which exhibit fluctuations in loss during the learning of dense embeddings. For $\text{KL}_\text{div}$, the relationship between $\textit{p}$ and $\textit{q}$ (see Section \ref{loss_function_3.1}) is investigated to show if $\textit{q}$  deviates from $\textit{p}$. This divergence is necessary because the data points in the target $\textit{p}$ require greater intra-cluster distance rather than high cohesion. We only consider those benchmarks using $\textit{p}$  and $\textit{q}$  distribution and $\text{KL}_\text{div}$ loss to align the comparison with TableDC. TableDC shows a high divergence (see Figure \ref{fig:loss_function}) of $\textit{q}$ from $\textit{p}$, which means that TableDC is more confident with the actual soft assignments $\textit{q}$ becoming sharper over time, and over highly packed $\textit{p}$, the $\textit{q}$ needs to minimize the cohesiveness.

\subsection{Self-supervision over different distance functions and similar kernels.}
\label{app:Self-supervision}

We evaluate the impact of difference distance measures and similarity kernels on the performance of TableDC, which utilizes a self-supervised module employing the Mahalanobis distance and Cauchy distribution to manage densely overlapping data effectively. We record the ARI and ACC (see Table \ref{tab:ablation:kernel}) of TableDC by keeping the Cauchy distribution as a similar kernel and replacing the Mahalanobis distance with Euclidean \cite{DBLP:conf/www/Bo0SZL020} and Cosine distances \cite{DBLP:journals/cacm/SaltonWY75}, each providing different geometric properties. The Euclidean distance measures the straight-line distance between data points and their centroids, and the Cosine distance evaluates the cosine of the angle between vectors, emphasizing orientation over magnitude. Similarly, we kept the Mahalanobis distance constant while varying the similarity kernels. We replaced the Cauchy distribution with the Student's t-distribution \cite{DBLP:conf/www/Bo0SZL020} (with an additional degree of freedom parameter that adjusts the distribution's kurtosis.) and the Normal distribution (that provides a standard Gaussian decay of similarity).

From Table \ref{tab:ablation:kernel}, we can observe that the choice of Mahalanobis distance combined with the Cauchy distribution appears optimal for all three problems. Euclidean distance with a Cauchy distribution performs relatively better in schema inference and domain discovery than cosine distance, showing that the straight-line distance between points is more reasonable than the angle when we have dense and overlapping clusters. With Student's t distribution, TableDC shows a high ARI for schema inference but significantly lower ACC than the Normal distribution. The higher ARI suggests that it can effectively group similar items into clusters, but the low ACC indicates that it failed to handle overlapping clusters and mismatches predicted labels against the actual labels.

\section{Conclusions}


This paper proposes a deep clustering algorithm for data cleaning and integration problems, and has demonstrated its applicability to schema inference, entity resolution and domain discovery. TableDC utilizes the Mahalonbis distance measure in its self-supervised module to optimize data distribution during training. The strength of TableDC is its ability to account for the covariance among different dimensions of the embeddings, often ignored by the conventional Euclidean distance. This ensures that the distance between data points remains consistent irrespective of the orientation of the data in the latent space. Our method enhances cluster separation in scenarios with overlapping features in the latent space by weighing dimensions based on the degree of their inter-dependencies. The experimental evaluation shows that TableDC consistently outperforms existing clustering algorithms and problem-specific  methods. 


\bibliographystyle{ACM-Reference-Format}
\bibliography{main}

\end{document}